\documentclass[12pt,thmsa]{article}
\usepackage{amsmath}
\usepackage{sw20nhj}

\input{tcilatex}
\begin{document}

\begin{titlepage}

\begin{center}
{\Large  \bf Quantum Effects of an Extra Compact
 Dimension on the Wave Function of the Universe}

\vspace{5mm}

\end{center}

\vspace{5 mm}

\begin{center}
{\bf Guowen Peng\footnote[1]{Department of Physics, 
Dalian University of Technology,
Dalian, 116024, People's  Republic of China.},
Hongya Liu\footnotemark[1]$^{,}$\footnote[2]
{\noindent Corresponding author, E-mail: hyliu@dlut.edu.cn.
}, and Qiuyang Zhang\footnotemark[1]}

\vspace{3mm}

\end{center}

\vspace{1cm}

\begin{center}
{\bf Abstract}
\end{center}
\baselineskip 23pt
We extend the direct quantum approach of the standard FRW cosmology 
from $4D$ to $5D$ and obtain a Hamiltonian formulation for a wave-like
 $5D$ FRW cosmology. Using a late-time approximation we isolate out a 
$y$-part from the full wave function of the $5D$ Universe. Then we find that 
the compactness of the fifth dimension $y$ yields a quantized spectrum for 
the momentum $P_{5}$ along the fifth dimension, and we show that the 
whole space-part of the wave function of the $5D$ Universe satisfies a 
two-dimensional Schr\"{o}dinger equation.

\vspace{1cm}
{\noindent \bf Key words:} Quantum cosmology; 
 Hamiltonian formulaiton.
\end{titlepage}\setcounter{page}{1}\baselineskip=23pt

\section{Introduction}

Quantum cosmology has received extensive studies (Wheeler et al. 1964;
Novello et al. 1995; Anchordoqui et al. 2000). Recently, a direct quantum approach was used to derive
Hamiltonian formulation of the standard Friedmann-Robertson-Walker (FRW)
cosmological models (Elbaz et al. 1997; Novello et al. 1996). The method is
based on assumptions of validity of the classical FRW equations and a
transformation from a pair of physical variables $(\rho ,\theta )$ (the
density of matter and the inverse of the Hubble radius), which describe the
dynamics of a spatially homogeneous and isotropic perfect fluid, to another
pair of canonical variables $(q,p)$ (roughly the radius of the Universe and
its rate of change). In this way, a canonical description of quantum FRW
cosmology was obtained. In this paper, we wish to extend this procedure to a 
$5D$ FRW cosmology and discuss quantum effects of the fifth compact
dimension on the wave function of the Universe.

The arrangement of this paper is as follows. In section 2, we study a class
of $5D$ wave-like cosmological solutions and derive Hamiltonian formulation
of the $5D$ FRW\ cosmology. In section 3, we use a late-time approximation
to separate the $5D$ wave function of the Universe and isolate out the part
of the fifth dimension. In section 4, we study quantization of the wave
function due to the compactness of the fifth dimension. Section 5 is a
conclusion.

\section{Hamiltonian Formulation of 5D Cosmology}

We consider a FRW type $5D$ metric

\begin{equation}
dS^{2}=-B^{2}(t,y)dt^{2}+A^{2}(t,y)(\frac{dr^{2}}{1-kr^{2}}+r^{2}d\Omega
^{2})+C^{2}(t,y)dy^{2}\;,  \tag{2.1}
\end{equation}
where $d\Omega ^{2}\equiv d\theta ^{2}+\sin ^{2}\theta d\varphi ^{2}$ and
the coordinates $x^{A}=(t,r\theta \varphi ,y)$ (Here and throughout this
paper, lowercase Greek letters run $0,123$ and uppercase Latin letters run $%
0,123,5$). Using the $4D$ part of the $5D$ metric (2.1) we can calculate all
the non-vanishing components of the $4D$ Einstein tensor $^{(4)}G_{\alpha
\beta },$ which are as follows: 
\begin{equation}
^{(4)}G_{0}^{0}=3\frac{\dot{A}^{2}}{A^{2}B^{2}}+\frac{3k}{A^{2}}\;, 
\tag{2.2}
\end{equation}
\begin{equation}
^{(4)}G_{1}^{1}=\;^{(4)}G_{2}^{2}=\;^{(4)}G_{3}^{3}=-2\frac{\ddot{A}}{AB^{2}}%
-\frac{\dot{A}^{2}}{A^{2}B^{2}}-\frac{k}{A^{2}}+2\frac{\dot{A}\dot{B}}{AB^{3}%
}\;.  \tag{2.3}
\end{equation}
where a dot denotes partial derivative with respect to time $t$.

It is known that solutions which are empty in $5D$ may have matter in $4D$
(Wesson 1999; Overduin \& Wesson 1997). Therefore, one can define an induced 
$4D$ energy-momentum tensor $T_{\alpha \beta \text{ }}$as 
\begin{equation}
T_{\alpha \beta }\equiv \;^{(4)}G_{\alpha \beta }\;.  \tag{2.4}
\end{equation}
It is found that this $T_{\alpha \beta }$ can take the form of a perfect
fluid, 
\begin{equation}
T_{\alpha \beta }=(\rho +P)u_{\alpha }u_{\beta }+Pg_{\alpha \beta }\;, 
\tag{2.5}
\end{equation}
where $\rho $ and $P$ are the energy density and pressure of the induced
matter, respectively.

Now we let the equation of state being 
\begin{equation}
P=\gamma \rho \;,  \tag{2.6}
\end{equation}
where $\gamma =0$ and $\gamma =\frac{1}{3}$ represent matter-dominated and
radiation-dominated eras, respectively. Then substituting (2.2), (2.3),
(2.5) and (2.6) in (2.4), we obtain the well-known Friedmann equation 
\begin{equation}
3\frac{\dot{A}^{2}}{A^{2}}=B^{2}\rho -\frac{3kB^{2}}{A^{2}}\;,  \tag{2.7}
\end{equation}
and the Raychaudhuri one 
\begin{equation}
3\frac{\ddot{A}}{A}-3\frac{\dot{A}\dot{B}}{AB}+\frac{B^{2}}{2}(1+3\gamma
)\rho =0\;.  \tag{2.8}
\end{equation}
The conservation law for the energy-momentum $T_{\alpha \beta }$ gives us
another equation 
\begin{equation}
\dot{\rho}+3(1+\gamma )\frac{\dot{A}}{A}\rho =0\;.  \tag{2.9}
\end{equation}

An exact $5D$ cosmological solution satisfying the empty $5D$ equations $%
R_{AB}=0$ is (Liu \& Wesson 1994) 
\begin{equation}
dS^2=-B^2(u)dt^2+A^2(u)(dr^2+r^2d\Omega ^2)+B^2(u)dy^2\;,  \tag{2.10}
\end{equation}
where $k$ is chosen to be zero,

\begin{equation}
\left. 
\begin{array}{c}
A(u)=(hu)^{\frac 1{2+3\gamma }}\;, \\ 
\;\;\;\;B(u)=(hu)^{-\frac{1+3\gamma }{2(2+3\gamma )}}\;,
\end{array}
\right.   \tag{2.11}
\end{equation}
and $u\equiv t-y.$ The energy density $\rho $ and pressure $P$ are 
\begin{equation}
\rho =\frac{3h^2}{(2+3\gamma )^2}A^{-3(1+\gamma )},\;\;P=\gamma \rho \;. 
\tag{2.12}
\end{equation}
This is a wave-like cosmological solution which can be interpreted as a
shock wave propagating along the fifth dimension (Wesson et al. 2000).

Now let us return to equations (2.7)-(2.9). There are three basic equations
from which we want to derive the Hamiltonian formulation. Note that since $%
A=A(u)$ and $u=t-y,$ we have $\dot{A}=\partial A/\partial t=dA/du.$
Therefore, by using a transformation 
\begin{equation}
\tilde{u}=\int B(u)du\;,  \tag{2.13}
\end{equation}
we can reduce equations (2.7)-(2.9) to 
\begin{equation}
\frac{3\stackrel{\ast }{A}^{2}}{A^{2}}=\rho \;,  \tag{2.14}
\end{equation}
\begin{equation}
\frac{3\stackrel{\ast \ast }{A}}{A}+\frac{1+3\gamma }{2}\rho =0\;, 
\tag{2.15}
\end{equation}
\begin{equation}
\stackrel{\ast }{\rho }+3(1+\gamma )\frac{\stackrel{\ast }{A}}{A}\rho =0\;, 
\tag{2.16}
\end{equation}
where an asterisk denotes derivative with respect to $\tilde{u}.$ These
three equations (2.14)-(2.16) are of the same forms as in the standard $4D$
FRW solutions algebraically. Therefore, we can follow the procedure given by
Elbaz et al. (Elbaz et al. 1997; see also Novello et al. 1996) to derive the
canonical formulation of the cosmology.

Firstly, we introduce an expansion parameter $\Theta ,\Theta $ $\equiv 3%
\frac{\stackrel{*}{A}}A$, which brings (2.14)-(2.16) to 
\begin{equation}
\frac 13\Theta ^2=\rho \;,  \tag{2.17}
\end{equation}
\begin{equation}
\stackrel{\ast }{\Theta }+\frac 13\Theta ^2+\frac 12(1+3\gamma )\rho =0\;, 
\tag{2.18}
\end{equation}
\begin{equation}
\stackrel{\ast }{\rho }+(1+\gamma )\rho \Theta =0\;.  \tag{2.19}
\end{equation}
Then we choose a new set of canonical variables $(q,p)$ with

\begin{equation}
\left. 
\begin{array}{c}
q=b\rho ^{-1/(3+3\gamma )}\;, \\ 
\;\;\;\;p=\frac 13b\Theta \rho ^{-1/(3+3\gamma )}\;,
\end{array}
\right.   \tag{2.20}
\end{equation}
where $b$ is an arbitrary constant. In this way, we arrive at the following
Hamiltonian 
\begin{equation}
\mathcal{H}(q,p)=\frac 12p^2-\frac 16b^{3(1+\gamma )}q^{-(1+3\gamma )}\;. 
\tag{2.21}
\end{equation}
This Hamiltonian describes a particle with momentum $p$ in a potential $%
V(q)=-\frac 16b^{3(1+\gamma )}q^{-(1+3\gamma )}$. Using relations
(2.17)-(2.21) we can verify that the two Hamilton equations $\stackrel{*}{q}%
=\partial \mathcal{H}/\partial p$ and $\stackrel{*}{p}=-\partial \mathcal{H}%
/\partial q$ hold. Be aware that here the time coordinate is not $t$ but $%
\tilde{u}$, that is, $\stackrel{*}{q}\equiv dq/d\tilde{u}$ and $\stackrel{*}{%
p}\equiv dp/d\tilde{u}$. This will make a difference between the two
canonical formulations of the $4D$ and $5D$ cosmologies as we will see in
next section.

\section{Wave Function and Late-Time Approximation}

From the Hamiltonian (2.21) we can employ the standard quantization
procedure to write the corresponding Schr\"{o}dinger equation. Because the
time coordinate in the Hamiltonian formulation is $\tilde{u}$ , so the
time-dependent wave function is of the form 
\begin{equation}
\Psi (q,\tilde{u})=e^{-i\tilde{E}\tilde{u}}\varphi (q)  \tag{3.1}
\end{equation}
The correspondence principle, $p\rightarrow -i\frac{\partial }{\partial q}$,
gives a stationary Schr\"{o}dinger equation for the stationary wave function 
$\varphi (q)$ as

\begin{equation}
\frac{1}{2}\frac{\partial ^{2}}{\partial q^{2}}\varphi (q)+[\tilde{E}+\frac{1%
}{6}b^{3(1+\gamma )}q^{-(1+3\gamma )}]\varphi (q)=0\;.  \tag{3.2}
\end{equation}
In (3.1) and (3.2) $\tilde{E}$ is the associated eigenvalue of the operator $%
\mathcal{\hat{H}}$, $\tilde{E}=<$ $\Psi (q,\tilde{u}$ $)|\mathcal{\hat{H}}%
|\Psi (q,\tilde{u}$ $)>$ . Quantized bound states with negative $\tilde{E}$
of the Schr\"{o}dinger equation (3.2) were studied (Novello et al. 1996;
Mongan 1999, 2000, 2001). Their results can also be used here.

Since $\tilde{u}$ is not the proper time of the cosmic fluid, we are not
sure if we can interpret $\tilde{E}$ as the energy. Therefore, we wish to
look for the relation between $\tilde{u}$ and the proper time. Now let us
consider the $5D$ metric (2.10), from which we see that the $2D$
line-element in the $t$ - $y$ plane can be written as 
\begin{equation}
ds^{2}=-B^{2}(u)dud(t+y)\;.  \tag{3.3}
\end{equation}
Thus by defining

\begin{equation}
U\equiv \int B^{2}(u)du\;,V\equiv t+y\;.  \tag{3.4}
\end{equation}
we get $ds^{2}=-dUdV.$ Then let 
\begin{equation}
U\equiv T-\lambda Y\;,V\equiv T+\lambda Y\;,  \tag{3.5}
\end{equation}
where $\lambda $ is a constant to be determined later, we get $%
ds^{2}=-dT^{2}+\lambda ^{2}dY^{2}.$ Therefore we find that $T$ is the proper
time. Substituting (2.11) in (2.13) and (3.4), we find 
\begin{equation}
h\tilde{u}=\frac{2(2+3\gamma )}{3(1+\gamma )}(hu)^{\frac{3(1+\gamma )}{%
2(2+3\gamma )}}\;,  \tag{3.6}
\end{equation}
\begin{equation}
hU=(2+3\gamma )(hu)^{\frac{1}{2+3\gamma }}\;.  \tag{3.7}
\end{equation}
From these two equations we obtain

\begin{equation}
\tilde{u}=W_{\gamma }U^{\frac{3(1+\gamma )}{2}}=W_{\gamma }(T-\lambda Y)^{%
\frac{3(1+\gamma )}{2}}\;,  \tag{3.8}
\end{equation}
where 
\begin{equation}
W_{\gamma }\equiv \frac{2}{3(1+\gamma )}(\frac{h}{2+3\lambda })^{\frac{%
1+3\gamma }{2}}  \tag{3.9}
\end{equation}

Now let us consider a short period in later times of the Universe. This
means that we choose $T=0$ as the beginning of the Universe (on the $Y=0$
hypersurface) and write $T=T_0+\tau $ with $|\tau |\ll T_0$. Then (3.8) gives
\begin{equation}
\left. 
\begin{array}{c}
\tilde{u}=W_\gamma T_0^{\frac{3(1+\gamma )}2}(1+\frac{\tau -\lambda Y}{T_0}%
)^{\frac{3(1+\gamma )}2} \\ 
\;\;\;\;\;\;\;\;\;\;\;\;\;\;\;\;\approx W_\gamma T_0^{\frac{3(1+\gamma )}2}+%
\frac{3(1+\gamma )}2T_0^{\frac{1+3\gamma }2}(\tau -\lambda Y)\;.
\end{array}
\right.   \tag{3.10}
\end{equation}

Using this relation we find that, up to a constant factor, the wave function 
$\Psi (q,\tilde{u})$ in (3.1) becomes 
\begin{equation}
\Psi (q,Y,\tau )\approx e^{-iE\tau }e^{iP_5Y}\varphi (q)\;.  \tag{3.11}
\end{equation}
where
\begin{equation}
\left. 
\begin{array}{c}
E\equiv \chi \tilde{E}\;, \\ 
\;P_5\equiv \chi \tilde{E}\lambda \;,
\end{array}
\right.   \tag{3.12}
\end{equation}

and 
\begin{equation}
\chi =\frac{3(1+\gamma )W_\gamma }2T_0^{(1+3\gamma )/2}=(\frac{hT_0}{%
2+3\gamma })^{\frac{1+3\gamma }2}\;.  \tag{3.13}
\end{equation}
Thus we successfully have obtained an approximate wave function $\Psi
(q,Y,\tau )$ as shown in (3.11), in which the three variables $\tau $, $q$
and $Y$ were separated.

\section{Quantization With a Compact Fifth Dimension}

The wave function $\Psi (q,Y,\tau )$ in (3.11) is in a form with three
separated variables $\tau ,q$ and $Y.$ Denote the $Y$-part of $\Psi $ as $%
\Phi (Y),$ $\Phi (Y)=e^{iP_{5}Y}$, then the momentum operator $\hat{P}_{5}=-i%
\frac{\partial }{\partial Y}$ gives 
\begin{equation}
\hat{P}_{5}\Phi =P_{5}\Phi \;.  \tag{4.1}
\end{equation}
So $P_{5}$ is the eigenvalue of the momentum along the fifth dimension.

Now we suppose the fifth dimension to be a circle with a radius $R$, i.e., $%
Y=R\phi $. So we have $\Phi =\Phi (\phi )=e^{iP_{5}R\phi }.$ The boundary
condition requires that $\Phi (\phi )$ must be periodic with period $2\pi $.
It follows that we must have

\begin{equation}
P_{5}=\frac{n}{R},\;\;n=0,\pm 1,\pm 2,...\;\text{.}  \tag{4.2}
\end{equation}
In this way, the fifth momentum $P_{5\text{ }}$and the corresponding wave
function $\Phi (\phi )$ are quantized.

The momentum operator $\hat{P}_{5\text{ }}=-i\frac \partial {\partial Y}$,
acting on Eq.(4.1) from the left hand side, gives 
\begin{equation}
\frac 12\frac{\partial ^2}{\partial Y^2}\Phi (Y)+\frac{n^2}{2R^2}\Phi
(Y)=0\;.  \tag{4.3}
\end{equation}
Let $\psi (q,Y)\equiv \varphi (q)\Phi (Y)$, then equations (3.2) and (4.3)
give 
\begin{equation}
\frac 12(\frac{\partial ^2}{\partial q^2}+\frac{\partial ^2}{\partial Y^2}%
)\psi (q,Y)+(\tilde{E}+\frac{n^2}{2R^2}+\frac 16b^{3(1+\gamma
)}q^{-(1+3\gamma )})\psi (q,Y)=0\;.  \tag{4.4}
\end{equation}
So we obtain a stationary two-dimensional Schr\"{o}dinger equation.

\section{Conclusion}

In this paper we have obtained a Hamiltonian formulation for a $5D$
cosmology. This Hamiltonian leads immediately to a stationary
one-dimensional Schr\"{o}dinger equation (3.2). Thus the corresponding full
wave function can describe quantized states of a $5D$ Universe. By using a
late-time approximation we have successfully separated the wave function
into three parts,$\Psi (q,Y,\tau )\approx e^{-iE\tau }e^{iP_{5}Y}\varphi (q)$
, corresponding to the proper time $\tau $, the variable $q$ (which
describes the radius of the Universe) and the fifth coordinate $Y$,
respectively. Notice that this wave function just valid in later times of the
Universe with $|\tau |\ll T_{0}$ , where $T_{0}$ is the age of the Universe
at that time. Then by assuming the compact fifth dimension to be a circle,
we obtain a quantized spectrum for the momentum $P_{5}$ in the fifth
direction. Be aware that the $q$-part of the wave function, $\varphi (q)$,
satisfies the one-dimensional Schr\"{o}dinger equation (3.2), for which
bound states were given (Novello et al. 1996). Using their results, as well
as the spectrum for $P_{5}$, we can obtain bound states for the whole
space-part of the wave function, $\psi (q,Y)=e^{iP_{5}Y}\varphi (q)$, which
satisfies the two-dimensional Schr\"{o}dinger equation (4.4). Further
studies are needed.

\noindent \textbf{Acknowledgments}

This work was supported by NSF of P. R. China under grant 19975007.

\end{document}